\documentstyle [10pt,emulateapj]{article}

\begin{document}

\title{ANISOTROPY IN THE COSMIC MICROWAVE BACKGROUND AT
DEGREE ANGULAR SCALES: \\ PYTHON V RESULTS}
\author{K. Coble\altaffilmark{1}, M. Dragovan\altaffilmark{1}, J. Kovac 
\altaffilmark{1}, N. W. Halverson\altaffilmark{1,8}, W. L. Holzapfel
\altaffilmark{1}, L. Knox\altaffilmark{1}, S. Dodelson\altaffilmark{2}, K.~Ganga~\altaffilmark{3,8}, D. Alvarez\altaffilmark{4}, J. B. Peterson\altaffilmark{4}, G. Griffin
\altaffilmark{4}, M. Newcomb\altaffilmark{4}, K. Miller\altaffilmark{5}, S. R. Platt
\altaffilmark{6}, G. Novak\altaffilmark{7}}

\altaffiltext{1}{Enrico Fermi Institute, University of Chicago, Chicago, IL 60637, coble@sealion.uchicago.edu}
\altaffiltext{2}{Fermilab, Batavia, IL 60510}
\altaffiltext{3}{IPAC, Pasadena, CA  91125}
\altaffiltext{4}{Carnegie-Mellon University, Pittsburgh, PA 15213}
\altaffiltext{5}{University of Colorado, Boulder, CO 80309-0440}
\altaffiltext{6}{University of Arizona, Tucson, AZ 85721}
\altaffiltext{7}{Northwestern University, Evanston, IL 60208-2900}
\altaffiltext{8}{California Institute of Technology, Pasadena, CA  91125}


\begin{abstract}

Observations of the microwave sky using the Python telescope 
in its fifth season of operation at the Amundsen-Scott South Pole Station
in Antarctica are presented.
The system consists of a 0.75 m off-axis telescope 
instrumented with a
HEMT amplifier-based radiometer having continuum sensitivity 
from 37-45 GHz in two frequency bands. With
a $0.91^{\circ} \times 1.02^{\circ} $ beam the instrument fully 
sampled 598 deg$^2$ of
sky, including fields measured during the previous
four seasons of Python observations. Interpreting the observed
fluctuations as anisotropy in the cosmic microwave background, we
place constraints on the angular power spectrum of fluctuations
in eight multipole bands up to $l \sim 260$. The observed spectrum
is consistent with both
the COBE experiment and previous Python results.
There is no significant contamination from known foregrounds.
The results show a discernible rise in the angular
power spectrum from large ($l \sim 40$) to small ($l \sim 200$)
angular scales.
The shape of the observed power spectrum is not a simple linear rise
but has a sharply increasing slope starting at $l \sim 150$.

\end{abstract}

\keywords{cosmic microwave background - cosmology: observations}


\section{INTRODUCTION}

Measurement of anisotropy in the Cosmic Microwave Background (CMB) 
directly probes conditions of the early universe.
Observations of the
angular power spectrum of CMB temperature fluctuations can be used to test
theories of structure formation and constrain cosmological
parameters. Results from the COBE satellite (Smoot et al. 1992)
tightly constrain the angular
power spectrum at the largest angular scales. Several experiments
have measured the angular power spectrum at degree angular
scales (e.g., Gaier et al. 1992, Schuster et al. 1993,
Gundersen et al. 1995, Lim et al. 1996, Platt et al. 1997, Cheng et al. 1997,  
Netterfield et al. 1997, Devlin et al. 1998, Herbig et al. 1998).
Collectively the data show a rise in
power towards smaller angular scales; individually no experiment covers
a wide range of angular scales from COBE scales to degree scales,
and most cover only small regions of the sky.
The dataset from the fifth Python observing season (hereafter PyV)
has sufficient sky coverage to probe the smallest scales
to which COBE was sensitive, while having a small enough beam to
detect the rise in angular
power at degree angular scales.

In its first four seasons the Python experiment detected significant
anisotropy in the CMB (Dragovan et al. 1994 (PyI), Ruhl et al. 1995 (PyII),
Platt et al. 1997 (PyIII), Kovac et al. 1999 (PyIV)). Observations from the
first three seasons were made at 90 GHz with a bolometer system
and a 4-point chop scan strategy, yielding CMB detections at
angular scales of $l \sim 90$ and $l \sim 170$. During the PyIV season
measurements were made using the same scan strategy with
a HEMT amplifier-based radiometer, confirming PyI-III detections 
in a 37-45 GHz frequency band. 

Observations were made from November 1996 through February 1997 in
the fifth Python observing season.
In order to increase the range of observed angular scales, a smoothly scanning
sampling scheme was implemented. As a result,
PyV is sensitive to the CMB angular power spectrum
from $l \sim$ 40 to $l \sim$ 260.


\section{INSTRUMENT}

The PyV measurements were made using the same receiver as the PyIV system
as described in Alvarez (1996) and Kovac et al. (1999).
The receiver consists of 
two focal-plane feeds, each with a single 37-45 GHz HEMT amplifier.
A diplexer splits each signal at $\sim$ 41 GHz before detection,
giving four data channels.  The analysis reported here
eventually combines signals from all four channels,
resulting in a thermal radiation
centroid $\nu_{c} = 40.3$ GHz and effective
passband $\Delta \nu = 5.7$ GHz for the PyV dataset.  

The receiver is mounted on a 0.75 m diameter
off-axis parabolic telescope (Dragovan et al. 1994),
which is surrounded by a large ground shield to
block stray radiation from the ground and Sun.
The beams corresponding to the two feeds observe the same
elevation and are separated by 2.80$^{\circ}$ on the sky. These beams
are scanned horizontally across the sky by a large rotating vertical flat
mirror, the chopper, at 5.1 Hz. 
The new scan strategy motivated two changes from the
instrument configuration described in Kovac et al. (1999):
the frequency response of the data system was
extended by switching to 100 Hz-rolloff antialiasing Bessel filters,
and the data recording rate was correspondingly
increased, to 652.8 samples/sec for each channel.


\section{CALIBRATION}

As in previous Python seasons, the primary DC calibration of the detectors
was derived using liquid nitrogen, liquid oxygen, and ambient
temperature thermal loads external to the receiver (Dragovan et al. 1994, Ruhl et al. 1995).
Load calibrations were performed approximately once per day,
and gains were found to be consistent over the
entire season to within $\pm$ 2\%, with no discernible trends.
Gain compression, which was a source of systematic
uncertainty in the calibration of
PyI-PyIII, is measured to be negligible for
the Python HEMT receiver.  Systematic uncertainty
in the DC load calibration is estimated to be $\pm$ 10\%.

Several efficiencies must be estimated to relate
the load calibrations to celestial response
in the main beam, which account for power losses
in the atmosphere, in the sidelobes,
and in the telescope, and they are calculated using data from skydips and 
from various beam measurements.
The resulting systematic calibration uncertainty of $^{+10\%}_{-4\%}$ is 
asymmetric, due to the fact that the individual losses
are small positive numbers and hence the errors in their
estimation follow skewed distributions (Kovac et al. 1999).

The dynamic response of the system was calculated from laboratory measurements
of the transfer functions for the AC coupling and
antialiasing filters in the data system,
and confirmed on the telescope by comparison of observations
made of the moon using normal and slow chopper speeds.
The response speed of the detectors is not a concern for this
calibration or for its uncertainty.
An appropriate response correction factor is applied
to each modulation of the data.
The uncertainty on these factors is small, and
is dominated by a $\pm$ 5\% systematic uncertainty on
their common normalization.

The overall uncertainty in the calibration of this
dataset is estimated to be $^{+15\%}_{-12\%}$.
Antenna temperature has been converted to units of $\delta T_{\rm CMB}$ throughout.


\section{OBSERVATIONS}
Two regions of sky were observed:  the PyV main
field, a $7.5^{\circ} \times 67.7^{\circ}$ region of sky centered at
$\alpha=23.18^{h}$, $\delta=-48.58^{\circ}$ (J2000) which includes 
fields measured during the previous four seasons of Python observations and
a $3.0^{\circ} \times 30.0^{\circ}$ region of sky centered
at $\alpha=3.00^{h}$,
$\delta=-62.01^{\circ}$ (J2000), which encompasses the region observed
with the ACME telescope (Gundersen et al. 1995).
The total sky coverage for the PyV regions is 598 deg$^2$,
greater sky coverage than previous degree-scale CMB experiments.
The combined absolute and
relative pointing uncertainty is estimated to be $0.15^{\circ}$, 
as determined by measurements of the
moon and the Carinae nebula ($\alpha=10.73h, \delta=-59.65 ^{\circ}$).
The PyV beam is well approximated by an asymmetric Gaussian
of FWHM $0.91^{+0.03}_{-0.01} \times 1.02^{+0.03}_{-0.01}$
degrees ($az \times el$).
The beam is determined from scans of the Carinae nebula and the Moon.
Both PyV regions are fully sampled 
with a grid spacing of 0.92$^{\circ}$ in elevation
and 2.5$^{\circ}$ in right ascension, corresponding to a distance
of 1.6$^{\circ}$ on the sky at a declination of $-50^{\circ}$.
The telescope is positioned on one of the fields
and the chopper smoothly scans the beams
in azimuth in a nearly triangular wave pattern. The chopper throw is
17$^{\circ}$ in azimuth, corresponding to 11$^{\circ}$ on the sky
at a declination of $-50^{\circ}$.
A total of 309 fields are observed in 31 sets of 5--17 fields.
Some of the fields are observed in more than one set.
There are 128 data samples for each detector channel
in a complete chopper cycle, and 164 chopper
cycles of data are taken of a given field before the telescope
is positioned on the next field in the set.
One data file consists of 164 chopper cycles for each field in the set.
Approximately 10 hours of good data (100 files) are taken of a set
of fields before the telescope moves on to the next set of fields.


\section{DATA REDUCTION}

After cutting 45\% of the data for weather and tracking errors,
389 hours of data remained for use in the CMB analysis.
The data are modulated using
\begin{equation}
M_{m}(\theta)=cos(m \pi \theta/\theta_c) \times
\left\{
\begin{array}{ll}
1 & m=1\\
H(\theta) & m=2 \ldots 8
\end{array}
\right.
\label{modeqn}
\end{equation}
where $m$ is the modulation number, $\theta$ is the
chopper angle, $\theta_c$ is
the extent of the chopper throw and
$H(\theta)~=~0.5(1~-~cos(2 \pi \theta/\theta_c))$ is a Hann window. (Fig.~1).
The $m=2 \ldots 8$ modulations are apodized with the Hann window 
in order to reduce the ringing of the window functions in $l$-space.
Data taken during the right- and left-going portions of the chopper
cycle are modulated separately, to allow for cross-checks of the data.
Sine modulations are not used in the analysis because they are anti-symmetric
and are thus sensitive to gradients on the sky.

\centerline{{\vbox{\epsfxsize=7.5cm\epsfbox{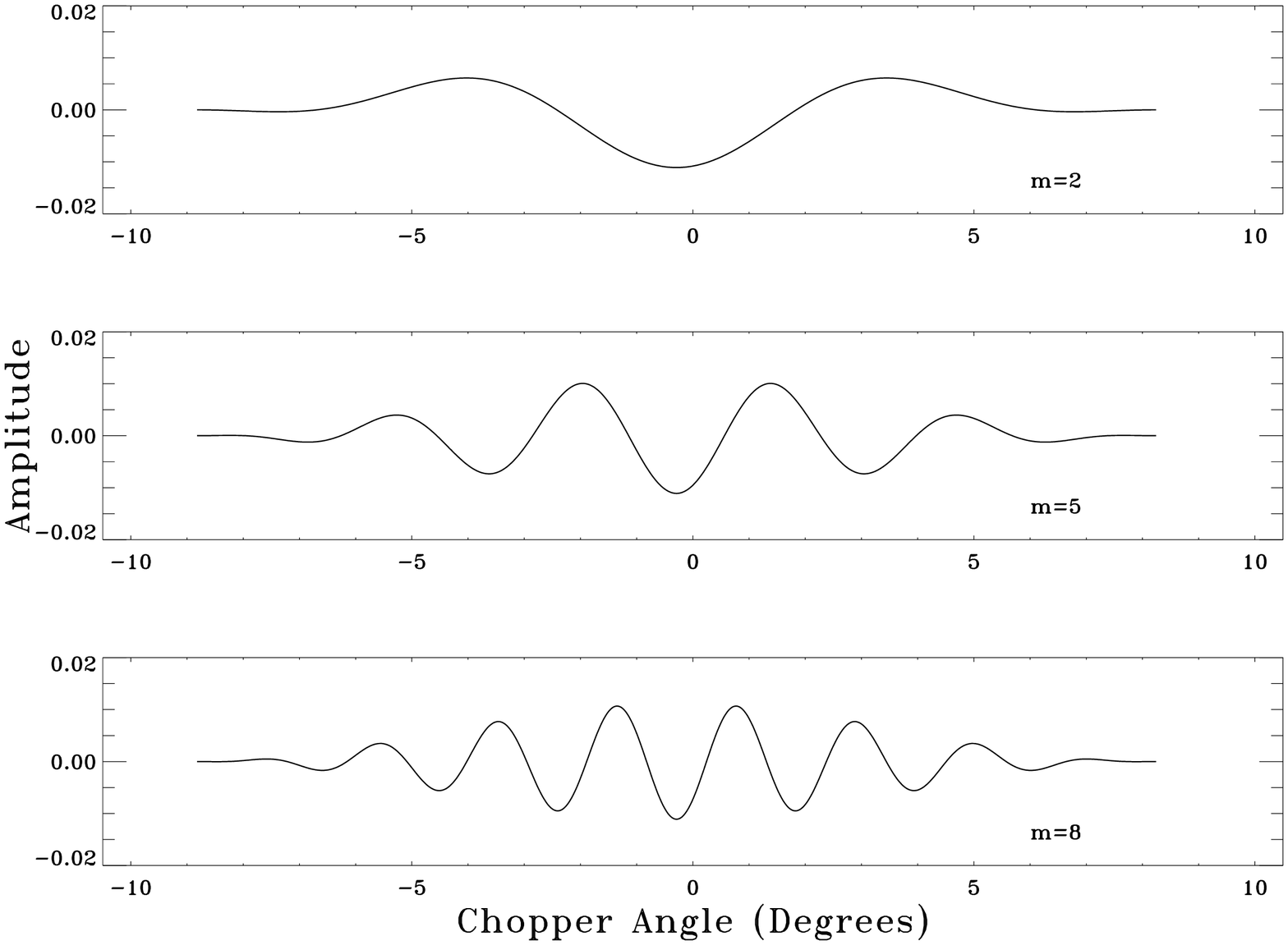}}}}
\figcaption{\footnotesize The data are modulated with cosines
which have been apodized by a Hann window. Three of the
modulations are shown.
\label{fig1}}

Data in a given file, field, channel, and modulation are co-added
over all chopper cycles.
A chopper synchronous offset, due to differential
spillover past the chopper, is removed from each data file
by subtracting the average of all of the fields
in a file. This is not just a DC offset; there is an offset removed
for each modulation (or equivalently, for each sample of the chopper
waveform) and channel. This subtraction must be accounted for by adding a term
to the covariance matrix (see section 6).
When the modulated data are binned in azimuth, a periodic signal due
to the 12 panels of the ground shield is evident, especially
on larger angular scales. The signal of period 30$^{\circ}$ is fit for
an amplitude and is subtracted. Removal of the ground shield offset
has less than 4\% effect on the final angular power spectrum (section 6)
in all modulations.

After the data have been modulated and offsets removed, the right and
left-going data, which have been properly phased, are co-added, as are
channels which observe the same points on the sky.
Since the two feeds observe different points on the sky they
cannot be co-added; the theoretical and noise covariances between them
are included in the likelihood analysis of the angular power spectrum.

The South Pole Station power nominally operates at a
frequency of 60 Hz, so there is
possible contamination at 60 Hz and its harmonics. However,
the chopper runs at 5.1 Hz, which is incommensurate with 60 Hz.

Our noise model assumes the covariance between fields taken with
different sets of files is negligible because of the chopper offset
removal and because of the long time (at least 10 hours) between measurements.
An analysis comparing the noise covariance estimated from data which
had not yet been co-added over all cycles and the
noise covariance estimated from data which had been co-added over all cycles
indicates that PyV noise is dominated by detector noise. However, the
long term drifts due to the atmosphere, which add to the variance
as well as induce small correlations between fields taken with the
same set of files, are important,
especially for power spectrum estimation (section 6).
The noise covariance between fields
taken with the same set of files is first
estimated by taking the usual covariance on the co-added data,
but because there were typically only 100 files
taken for each field, the sample
variance on the noise estimate is $\sim (100)^{1/2}$, or 10\%, which will
severely bias estimates of band power. To obtain
a better estimate of the noise, we averaged the
variances for each set of files and then scaled the off-diagonal
elements of the covariance
to the average variance in a given set based
on a model derived from the entire PyV data set.

Several self-consistency checks were performed on the data set, using
our best estimate of the noise covariance.
The $\chi^2={\bf d}^tC^{-1}{\bf d}$, where ${\bf d}$ is the data vector and
$C=C^T+C^N+C^C$ is the total covariance (see section 6), is consistent with
its expected value (the number of degrees of freedom) in
all modulations.
The value of the probability enhancement factor, $\beta$,
(Knox et al. 1998) between data from the two feeds falls within
the expected range for all modulations. Finally, the data
set was transformed into the
signal-to-noise eigenmode basis. In that basis,
$C^{-1/2}{\bf d}$ should be Gaussian distributed with
$\sigma=1$ and a total area equal to the number of degrees of freedom.
Histograms of $C^{-1/2}{\bf d}$ are consistent with such a Gaussian
distribution for all of the modulations.
We performed these same tests using the preliminary noise covariance,
which suffers from sample variance, and found that the
consistency tests failed.
The tests indicate that the data set used is self consistent
and that our best estimate of the noise covariance is indeed
a good model for the noise.


\section{DATA ANALYSIS AND RESULTS}

CMB angular power is usually expressed in terms
of angular multipoles, $C_{l}$. A flat power spectrum is one for
which ${\cal C} \equiv (l(l+1)C_{l}/2\pi)$ is constant.
For each of the eight modulations, we compute 
the likelihood (${\cal L}$) as a function of ${\cal C}$.
The theoretical covariance matrix, $C^T$, needed
to compute ${\cal L}$ depends on the amplitude of ${\cal C}$
and the experimental window functions.
The window functions are a measure of experimental sensitivity as a function
of angular scale $l$ (Fig. 2). They are generated
from the experimental beam map,
modulations, and observing strategy and are given by
\begin{equation}
W_{lij} = {(1/2\pi)} \int d\phi
e^{-i {\bf k} \cdot ({\bf x_{i}} - {\bf x_{j}})}
\tilde B({\bf k}) \tilde B^{*}({\bf k})
\label{wlij}
\end{equation}
where $\tilde B({\bf k})$ is the Fourier transform of the
beam map for the given modulation, ${\bf x_{i}}$
the position of field $i$, and ${\bf k} = l (\cos\phi,\sin\phi)$.
These functions are computed for all pairs of fields and channels.

\centerline{{\vbox{\epsfxsize=8.0cm\epsfbox{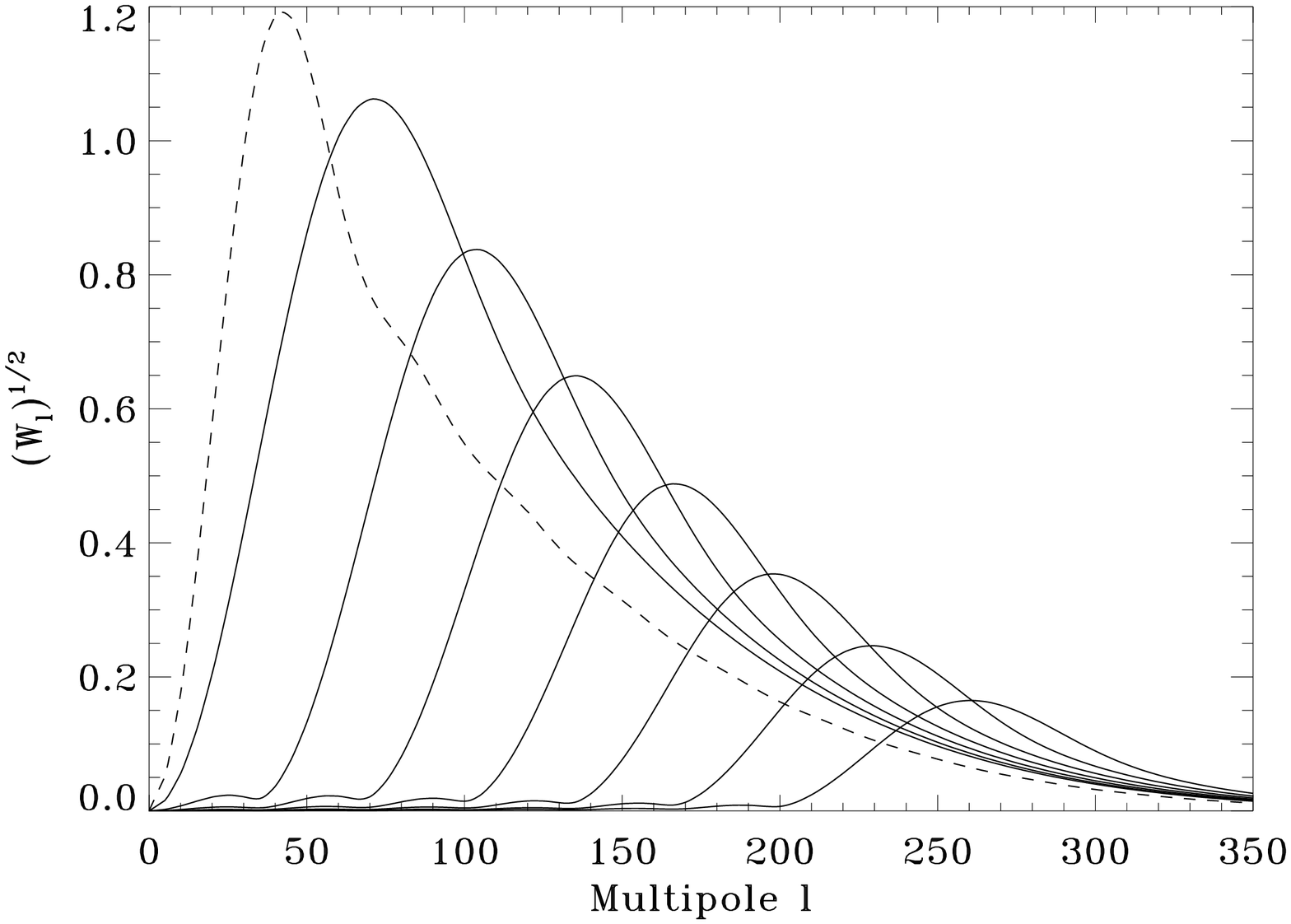}}}}
\figcaption{\footnotesize Diagonal window functions. The unapodized
cosine modulation is plotted with a dashed line and the apodized
cosine modulations are plotted with solid lines.
\label{fig2}}

The subtraction of the chopper synchronous offset can be
accounted for with an additional term in the covariance
matrix (e.g., \cite{lloyd}) and in this case is given by
$C^{C}=(K/N_{\rm files})$
for fields taken with the same set of files and zero
for fields not taken with the same
set of files. 
$N_{\rm files}$ is the number of files which observed the fields.
Taking $K$ to be large ensures that we have 
no sensitivity to modes of the data
that could have come from the chopper.

The CMB power spectrum is shown in Fig. 3 and is given in Table 1.
The band powers are calculated for each modulation
separately. This neglects the correlations between modulations.
Work is in progress to fit the amplitudes in
a series of bands to all of the modulations simultaneously.
The calibration uncertainty 
allows all band powers to shift by the same amount (i.e. the calibration
errors are correlated). Given the uncertainty
in the beam size of approximately $0.015^\circ$,
the band power for a given modulation can move roughly by a factor of
$\exp ( \pm l(0.425)(0.015)(\pi/180) )$, only a $3\%$
effect at $l=200$. 
Flat band powers calculated from the subset of PyV data in the PyIII
region of sky are consistent with the PyIII flat band powers
to within the uncertainties.
Given the $^{+15\%}_{-12\%}$ calibration uncertainty
and the $3 \mu$K statistical
uncertainty, the lowest $l$ PyV modulation agrees with the 
the smallest scale COBE measurements.

\bigskip
\bigskip
\bigskip

\begin{center}
{Table 1\\ \footnotesize Results of likelihood analysis. Band powers are in $\mu K$.}
{\scriptsize
\begin{tabular}{|ccc|}
\tableline
{\bf mode} & {\bf $l_{e}$}    &  {\bf $(l(l+1)C_{l}/2\pi)^{1/2}$} \\
\tableline
1 &  $50^{+44}_{-29}$   &  $23^{+3}_{-3}$ \\
2 &  $74^{+56}_{-39}$   &  $26^{+4}_{-4}$ \\
3 &  $108^{+49}_{-41}$  &  $31^{+5}_{-4}$ \\ 
4 &  $140^{+45}_{-41}$  &  $28^{+8}_{-9}$\\
5 &  $172^{+43}_{-40}$  &  $54^{+10}_{-11}$ \\
6 &  $203^{+41}_{-39}$  &  $96^{+15}_{-15}$\\
7 &  $233^{+40}_{-38}$  &  $91^{+32}_{-38}$\\
8 &  $264^{+39}_{-37}$  &  $0^{+91}_{-0}$  \\
\tableline
\end{tabular}}
\end{center}


\section{FOREGROUNDS}
The PyV data are cross-correlated with several foreground templates in
order to set limits on possible foreground contamination. The
templates used are the Schlegel et al. (1998) 100 micron dust map,
which is based on IRAS and DIRBE maps,
the Haslam et al. (1974) 408 MHz survey (synchrotron),
and the PMN survey (point sources).
Each foreground template map is smoothed to
PyV resolution, pixelized and
modulated according to the PyV observation scheme.

Two templates are created for the PMN survey.
We call one PMN, which is converted to
$\delta T_{\rm CMB}$ using the spectral indices given in the survey. The
other we call PMN0, which is
converted to a flux at 40 GHz assuming
a flat spectrum extrapolated from the flux measurement at 4.85 GHz.
The assumption of a flat spectrum is  
conservative in that it is likely to over-estimate the flux
at 40 GHz. Neither case is correct, since spectral indices have not
been measured for all of the sources, in which case a flat spectrum is
assumed, but we do know that some of them are not flat, so a flat
spectrum will be inappropriate. The two cases cover a
reasonable range of possibilities.

For each modulation and foreground template, a
correlation coefficient and uncertainty are calculated
following de~Oliveira-Costa et al. (1997).
A weighted mean and uncertainty over all of the modulations
for each foreground are given in Table 2.
In all cases there is no clear detection of foregrounds. 
The RMS of each modulation of each foreground was calculated
and then multiplied by the corresponding $1 \sigma$
error bar in Table 2 in order to estimate an upper limit on the
foreground contribution to CMB band power.
The limits on contributions from
foregrounds are given in Table 3 and are
at least $\sim 10 \times$ smaller than the measured CMB bandpowers.

\bigskip
\bigskip
\bigskip

\begin{center}
{Table 2\\ \footnotesize Correlation coefficients and
uncertainties for weighted means.}

{\scriptsize
\begin{tabular}{|cccc|}
\tableline
Dust  & Haslam  & PMN  & PMN0 \\
$\mu K (MJy/sr)^{-1}$ & $\mu K / K$ & $\mu K / \mu K$ & $\mu K(MJy/sr)^{-1}$\\
\tableline
-3 $\pm$ 18  &  -2.0 $\pm$ 2.6  &  0.012 $\pm$ 0.024  &  195 $\pm$ 385\\
\tableline
\end{tabular}}
\end{center}

\begin{center}
{Table 3\\ \footnotesize Upper limits on foreground contribution. All
units are $\mu K$.}

{\scriptsize
\begin{tabular}{|ccccc|}
\tableline
Mode & Dust  &  Haslam & PMN  & PMN0\\
\tableline
1 &  1.0  & 0.5    & 0.3  & 0.3 \\
2 &  1.0  & 0.6    & 0.6  & 0.6 \\
3 &  1.1  & 0.8    & 1.0  & 1.0 \\
4 &  1.1  & 1.0    & 1.1  & 1.1 \\
5 &  1.9  & 2.7    & 2.7  & 2.7 \\
6 &  2.7  & 6.7    & 5.4  & 5.4 \\
7 &  2.0  & 6.9    & 5.9  & 5.9 \\
8 &  1.6  & 8.2    & 6.6  & 6.6 \\
\tableline
\end{tabular}}
\end{center}

If the diffuse morphology of the sky is not
constant as a function of wavelength, then these
templates do not reveal all of the foreground
contamination and more could be hidden in the PyV data.
A combined analysis of PyIII at 90 GHz and PyV at 40 GHz would
constrain the foreground contamination further.
When the data themselves are analyzed as a whole, rather than
separately for each modulation, the foreground analysis will
also be done on the data set as a whole. This will take into
account the cross-modulation correlations in the data.


\section{CONCLUSIONS}

The PyV experiment fully samples 598 deg$^2$ of the microwave sky
and constrains the CMB angular power spectrum in the angular scale
range $40 \lesssim l \lesssim 260$.
The measurements pass internal consistency checks,
show little contamination from foreground radiation, and
are consistent with previous Python and COBE results.
The observed angular power increases from larger
to smaller angular scales, with a sharply increasing slope starting
at $l \sim$ 150. 


\acknowledgments

We would like to thank Bharat Ratra for helpful conversations and for
aiding us in checking our window functions.
This work was supported by the James S. McDonnell Foundation, PYI grant
NSF AST 90-57089, and the NSF under a cooperative agreement with the
Center for Astrophysical Research in Antarctica (CARA), grant NSF OPP
89-20223. CARA is an NSF Science and Technology Center.
The work of SD was supported by NASA Grant NAG 5-7092 (in addition to the DOE).
KG acknowledges support from NASA ADP grant NASA-1260. KC is supported by
NASA NGT 5-19.


\newpage

\rightline{\scriptsize Coble et al. 1999}
\centerline{\large Python V}
\centerline{{\vbox{\epsfxsize=20cm\epsfbox{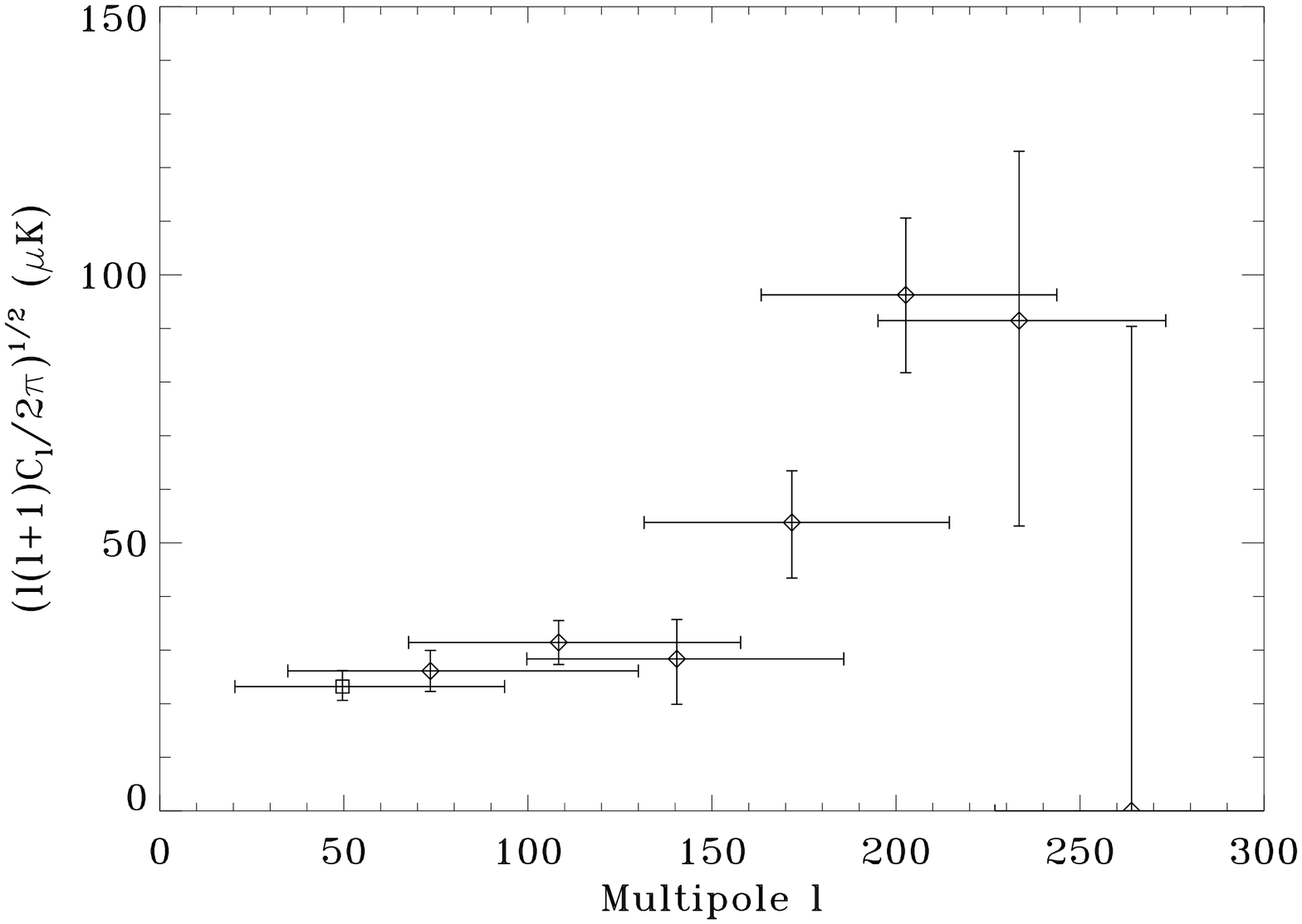}}}}
\figcaption{\footnotesize Flat band power, $(l(l+1)C_{l}/2\pi)^{1/2}$, vs.
multipole $l$ for all of the modulations.
The detections have $1 \sigma$ error bars and the upper limit has $2 \sigma$
error bars.
The unapodized cosine modulation is plotted with an open square
and the apodized cosine modulations are plotted
with diamonds. The error bars include
statistical uncertainties only and do not include uncertainties in the
calibration or beam size.
The $l$ range of each modulation is determined by the half-maximum points of
$(W_l)^{1/2}$. Low $l$ values correspond to large angular
scales and high $l$ values correspond to small angular scales. CMB power is
clearly rising from low to high $l$ up to the sensitivity cutoff of PyV.
\label{fig3}}


\begin{thebibliography}{}

\bibitem[(Alvarez 1996)]{alv} Alvarez, D. 1996, Ph.d. thesis, Princeton Univ.

\bibitem[(Bond et al. 1998)]{lloyd} Bond, J. R., Jaffe, A., and Knox, L. 1998,
\prd, 57, 2117

\bibitem[(Cheng et al. 1997)]{msam} Cheng, E. S., et al. 1997, \apj, 488, L59

\bibitem[(de Oliveira-Costa et al. 1997)]{angelica} de Oliveira-Costa, A., et al. 1997, \apj, 482, 17

\bibitem[(Devlin et al. 1998)]{devlin} Devlin, M., et al. 1998, \apjl, 509, L69

\bibitem[(PyI)]{py1} Dragovan, M., et al. 1994 \apjl, 427, L67 (PyI)

\bibitem[Gaier et al. 1992]{gaier} Gaier, T. et al. 1992, \apjl, 398, L1

\bibitem[(Gundersen et al. 1995)]{jg} Gundersen, J. O., et al. 1995, \apjl, 443, L57

\bibitem[(Haslam et al. 1974)]{haslam} Haslam, C.G.T. et al. 1974, \aaps, 13, 369

\bibitem[(Herbig et al. 1998)]{herbig} Herbig, T., et al. 1998, \apjl, 509, L73

\bibitem[(PyIV)]{py4} Kovac, J., et al. 1999, in preparation (PyIV)

\bibitem[(Knox et al. 1998)]{beta} Knox, L., et al. 1998, \prd, 58, 083004

\bibitem[(Lim et al.)]{max} Lim, M. A., et al. 1996, \apjl, 469, L69

\bibitem[(Netterfield et al.)]{sk} Netterfield, C. B., et al. 1997, \apj, 474, 47

\bibitem[(PyIII)]{py3} Platt, S. R., et al. 1997, \apjl, 475, L1 (PyIII)

\bibitem[(PyII)]{py2} Ruhl, J. E., et al. 1995, \apjl, 453, L1 (PyII)

\bibitem[Schuster et al. 1993]{schuster} Schuster, J. et al. 1993, \apj,
412, L47 

\bibitem[(Schlegel et al. 1998)]{dust} Schlegel, D. J., Finkbeiner, D. P. 
and Davis, M. 1998, \apj, 500, 525

\bibitem[(Smoot et al. 1992)]{cobe} Smoot, G. F., et al. 1992, \apjl, 396, L1


\end{thebibliography}
\end{document}